\documentstyle[11pt,newpasp,twoside,epsf]{article}
\newcommand{\mum}{${\rm \mu m}$}
 
\newcommand{\waven}{${\rm cm^{-1}}$}
\newcommand{\thirteenco}{${\rm ^{13}CO}$}

\def\edcomment#1{\iffalse\marginpar{\raggedright\sl#1\/}\else\relax\fi}
\reversemarginpar

\begin{document}
\title{Interstellar Ices}
\author{A. C. A. Boogert}
\affil{California Institute of Technology, MS 105-24, Pasadena, CA
91125; acab@astro.caltech.edu}
\author{P. Ehrenfreund}
\affil{Leiden Observatory, P. O. Box 9513, 2300 RA Leiden, the
Netherlands}

\begin{abstract}
Currently $\sim$36 different absorption bands have been detected in
the infrared spectra of cold, dense interstellar and circumstellar
environments.  These are attributed to the vibrational transitions of
$\sim$17 different molecules frozen on dust grains. We review
identification issues and summarize the techniques required to extract
information on the physical and chemical evolution of these ices. Both
laboratory simulations and line of sight studies are essential.
Examples are given for ice bands observed toward high mass protostars,
fields stars and recent work on ices in disks surrounding low mass
protostars. A number of clear trends have emerged in recent years.
One prominent ice component consists of an intimate mixture between
H$_2$O, CH$_3$OH and CO$_2$ molecules. Apparently a stable balance
exists between low temperature hydrogenation and oxidation reactions
on grain surfaces.  In contrast, an equally prominent ice component,
consisting almost entirely of CO, must have accreted directly from the
gas phase. Thermal processing, i.e. evaporation and crystallization,
proves to be readily traceable in both these ice components.  The
spectroscopic signatures of energetic processing by cosmic rays and
high energy photons from nearby protostars are weaker and not as well
understood. A fundamental limitation in detecting complex,
energetically produced (and also some simple) species is blending of
weak features in the spectra of protostars.  Sophisticated techniques
are required to extract information from blended features. We conclude
with a summary of key goals for future research and prospects for
observations of ices using future instrumentation, including
SIRTF/IRS.
\end{abstract}

\section{Introduction}

The infrared spectra of continuum emitting sources (stars, protostars,
evolved stars, active galactic nuclei) extincted by large columns of
cold dust in molecular clouds, protostellar envelopes, disks, shells,
or circumnuclear rings, show a wealth of absorption features. Most of
these are attributed to the vibrational modes of volatile solid state
molecules ($T_{\rm subl}\leq 100$ K).  Over half a century ago (Bates
\& Spitzer 1951) it was realized that the formation of 'ices' is
catalyzed by the cold surfaces of silicate and carbon-based dust
particles originating from the outflows of novae, supernovae, and
stars on the asymptotic giant branch (see reviews by Patzer et al.,
Jura et al., and Clayton et al. in these proceedings). When such
particles enter the dense interstellar medium, atoms and molecules
strike and stick to the surfaces of these cold dust grains. At average
molecular cloud densities ($\sim 10^4~{\rm cm^{-3}}$), atoms and
molecules accrete from the gas at the rate of about one per day (e.g.,
Tielens \& Allamandola 1987). The sticking coefficients for most of
these species at 10 K are close to unity with the exception of He and
H$_2$. Atoms (such as H, D, C, O) are mobile on the surface and are
able to find a compatible reaction partner.  Hydrogen atoms migrate
rapidly across the surface by quantum-mechanical tunneling (Manico et
al. 2001). Heavy atoms (e.g., C and O) can diffuse by thermal hopping
(e.g., Tielens \& Allamandola 1987). Efficient adsorption and
diffusion of species allows chemical reactions on the surface to occur
and to form simple and complex molecular species by, for example,
exothermic hydrogenation reactions (e.g. Tielens \& Hagen 1982;
Charnley et al. 1992).

\begin{figure}
\plotfiddle{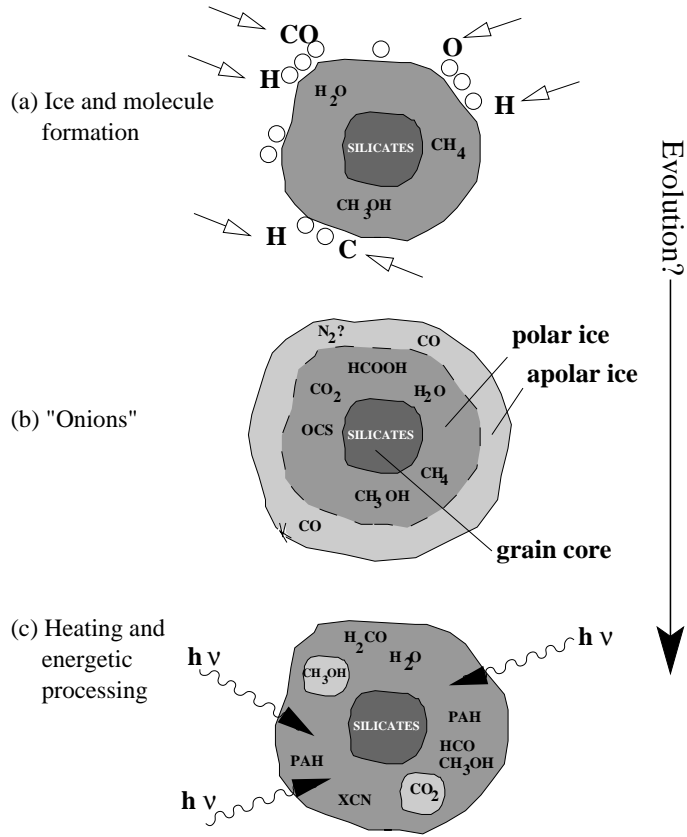}{310pt}{0}{65}{65}{-140pt}{0pt}
\caption{A grain in space. Possible formation and evolution processes
of the icy mantles are indicated.  Interstellar grains are
schematically represented by spheres, although this is likely
unrealistic (e.g. \S 5).}
\end{figure}

At high atomic abundances, for example early in the cloud life time or
at cloud edges, an H$_2$O--rich ice mantle is thus rapidly
formed. Since hydrogenated species like H$_2$O, and also CH$_3$OH and
NH$_3$ have large dipole moments, these types of ices, including trace
species diluted in them, are referred to as `polar' ices in the
literature (Fig.~1a).  At low atomic abundances and low temperatures,
i.e. deeper in the cloud, the accreting gas is primarily molecular and
an `apolar' ice, rich in CO and possibly N$_2$ or O$_2$ is
formed. Depending on the ice formation history (e.g.  increasing
densities and molecular abundances as a function of time, or migration
of grains in a turbulent cloud) apolar and polar ices may be present
on the same grain in an onion-like structure (Fig. 1b), or at
different locations along the same line of sight.  Polar and apolar
ices can be distinguished spectroscopically (different band profiles),
and can help trace the thermal and chemical evolution of ices, clouds,
and comets (apolar ices evaporate at $\sim$20 K and polar ices at
$\sim$90 K).

The high sticking coefficients would suggest that accretion leads to
extreme depletions in the molecular gas on time-scales of about a few
10$^5$ years (e.g., Brown \& Charnley 1990). However, astronomical
observations of molecular gas in the dense interstellar medium
indicate also efficient desorption of material from the grain surface.
Several desorption mechanisms have been proposed including evaporation
following grain heating by cosmic rays, mantle explosions, and
ejection upon molecule formation (see Willacy \& Millar 1998).  As
dense clouds are often regions of star formation, dynamical events
also have been proposed for removing and reforming the molecular ices
within the accretion time-scale; these include sputtering in
low-velocity shock waves (Bergin et al. 1998; Charnley et al. 2001)
and grain-grain collisions (Markwick et al.  2000; Dickens et
al. 2001).

The interaction between the interstellar gas and grains and the
prevailing conditions in the line of sight are dominant factors in
grain evolution. After icy grain mantles are formed through gas
condensation and surface chemistry, the material can be hit by
energetic photons and cosmic rays, which triggers dissociation and
recombination mechanisms, and thermal processing which may result in
crystallization, clathrate formation, polymerization and acid base
chemistry (Fig. 1c; Ehrenfreund \& Schutte 2000).  

Determining the importance of the above-mentioned ice formation and
evolution scenarios is a key topic in astrophysics and relates
directly to studies of the origin and evolution of solar system ices
(e.g. Mumma et al. 2003). In this review we discuss the observational
status of interstellar ices. We present an up-to-date list of detected
absorption features and discuss their identification and uncertainties
in \S 2. We discuss how information on ice origin and evolution can be
derived from observations by comparison with laboratory simulations
and line of sight characteristics (\S 3). Detailed band profile
studies of ices in different environments and the significance of
thermal processing are discussed in \S 4.  In \S 5 we touch upon the
importance of observing isotopes in the ices and deriving the shape of
interstellar grains. Subsequently we discuss the observational
evidence for evolution of interstellar ices from a chemical
perspective (\S 6).  Finally, we summarize outstanding issues and what
can be expected from observations of interstellar ices with future
instrumentation (\S 7).

\begin{figure}
\plotfiddle{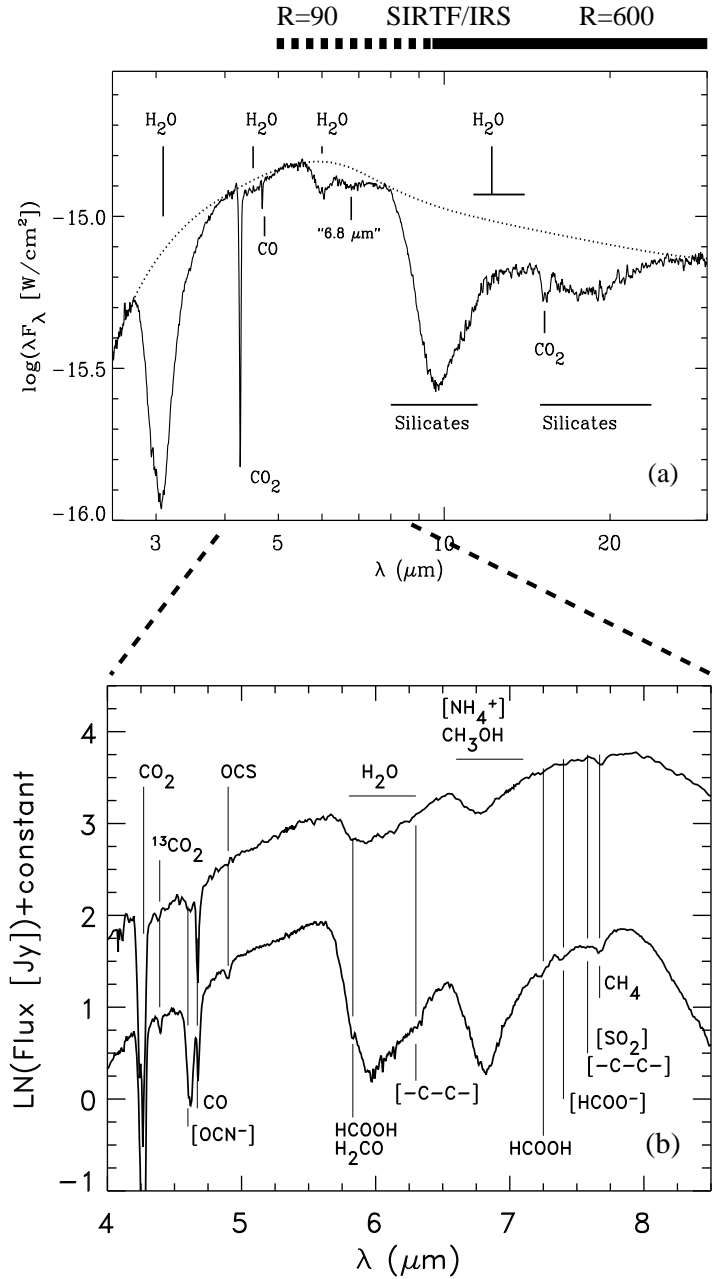}{490pt}{0}{90}{90}{-170pt}{0pt}
\caption{Mid-infrared ISO/SWS spectra of protostars showing the
richness of ice absorption features. (a) The low mass protostar Elias
29 in $\rho$ Oph at a spectral resolution of $R$=400 (Boogert et
al. 2000a).  The bar on top indicates the wavelength coverage and
spectral resolution of the SIRTF/IRS instrument (\S 7). (b) The rich
mid-infrared spectra of the massive protostars NGC 7538 : IRS9 (top)
and W 33A (bottom). Data taken from Whittet et al. (1996) and Gibb et
al. (2000). Identifications labeled in square brackets are uncertain.}
\end{figure}

\section{Inventory of Ice Absorption Features}

A large number of infrared absorption features attributed to ices has
been detected. Complex mid-infrared absorption spectra have been
observed toward low and high mass protostars (Fig. 2) and field stars.
These features have depths that vary widely from source to source, but
often they absorb a significant fraction of the continuum photons. In
fact, the observed integrated absorption optical depth (in wavenumber
space) of ices in the 2-30 \mum\ range can easily exceed that of
silicates. For example, in the line of sight towards Elias 29, a
protostar with moderately deep ice bands, the integrated optical depth
of ices is $\sim$1000 \waven, or 3 times that of silicates!

\begin{figure}[t]
\plotfiddle{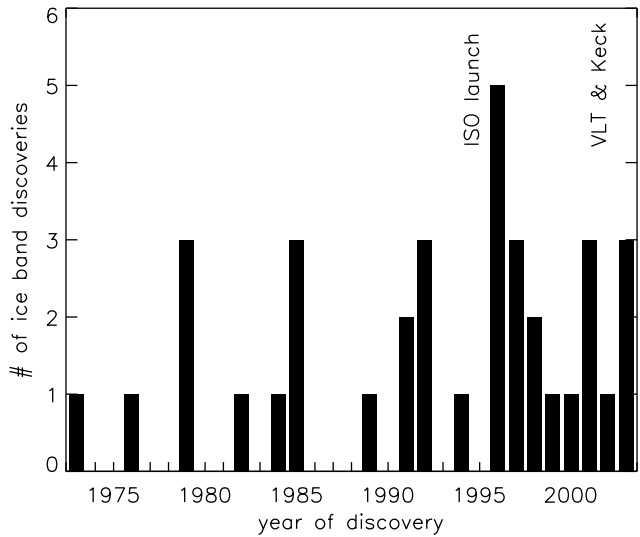}{180pt}{90}{50}{50}{200pt}{-20pt}
\caption{Detection of ice features in the last 30 years.  The
availability of the space based ISO spectrometer and 2-5 \mum\
spectrometers at large telescopes mark a substantial increase in band
detection.}
\end{figure}

The first ice band, at 3.07 \mum\ from the O--H stretch mode of
H$_2$O, was discovered in 1973 in the BN object (Gillett and Forrest
1973).  Since then gradual improvements in detector size, sensitivity
and telescope size have led to the discovery of roughly one feature
per year, with boosts initiated by the ISO satellite and recently by
high resolution spectrometers at VLT and Keck (Fig. 3). The latest
detections, including the 2.27 \mum\ band from a CH$_3$OH combination
mode (Taban et al. 2003), a shallow band at 4.60 \mum\ possibly
related to CO (Pontoppidan et al. 2003a), and the 3.32 \mum\ band from
the C--H stretch mode of CH$_4$ (Boogert et al. 2003), show that
improvements in instrumentation keep driving this area of
astrophysics.

\begin{table}
\caption{Interstellar ice absorption features$^{\rm a}$}
\center
\begin{tabular}{p{8cm}l}
%\noalign{\smallskip} 
\tableline
\noalign{\smallskip} 
$\lambda_{\rm peak}^{\rm b}$ & Molecule$^{\rm c, d}$     \\
\mum                          &              \\
\noalign{\smallskip} 
\tableline
\noalign{\smallskip} 
\tableline
\noalign{\smallskip} 
4.67 & CO \\
4.78 & \thirteenco\  \\
2.96, 3.07, 3.2--3.7, 4.5, 6.0, 12, 44 & H$_2$O \\
2.70, 2.78, 4.27, 15.2 & CO$_2$ \\
4.38 & \thirteenco$_2$ \\
4.92 & OCS \\
7.60 & [SO$_2$] \\
2.96, 3.2--3.7, 3.47, 9.01 & [NH$_3$] \\
7.41 & [HCOO$^-$] \\
5.83 (narrow) & H$_2$CO \\
5.83 (broad), 7.25 & HCOOH \\
3.32, 7.67 & CH$_4$ \\
3.25, 3.47, 6.85 & [NH$_4^+$], [PAH, -C-H] \\
2.27, 3.54, 3.85, 3.94, 4.1,  6.85, 8.9, 9.7 & CH$_3$OH \\
4.62 & [OCN$^-$] \\
6.25, 7.60 & [PAH, C-C] \\
\noalign{\smallskip} 
\tableline
\tableline
\noalign{\smallskip} 
\multicolumn{2}{p{13cm}}{a: for more details see
http://www.astro.caltech.edu/$\sim$acab/icefeatures.html}\\
\multicolumn{2}{p{13cm}}{b: features listed more than once
have multiple possible identifications.}\\
\multicolumn{2}{p{13cm}}{c: uncertain assignments
are enclosed in square brackets.}\\
\multicolumn{2}{p{13cm}}{d: see Fig. 4 for abundances.}\\
\end{tabular}
\end{table}

Table 1 lists all 36 absorption features that have been attributed to
ices so far, along with identifications, or tentative
identifications\footnote{A more detailed and regularly updated list of
ice features, including references can be found at
{http://www.astro.caltech.edu/$\sim$acab/icefeatures.html}}.  About
half the listed molecules are securely identified. Simple molecules
with strong absorption bands from multiple vibrational modes are most
securely identified (H$_2$O, CH$_3$OH, CO$_2$). To this can be added
the species with isotope detections (CO, CO$_2$). The identifications
of some simple molecules are however still controversial (NH$_3$), as
is the identification of a strong absorption band (6.85 \mum). We will
discuss these in some detail (\S\S 2.2 and 2.3), and summarize issues
on the weakest ice components (\S 2.4).

\subsection{Interstellar Ice Abundances}

H$_2$O is the most common interstellar ice component at an abundance
of typically 1$\times 10^{-4}$ with respect to the total H column
density ($N$[H]+2$N$[H$_2$]; e.g. Tielens et al. 1991). As summarized
in Fig. 4, CO$_2$ is the second most common ice component with an
abundance of on average 17\% with respect to H$_2$O ice over many
lines of sight (Gerakines et al. 1999). Although three intriguing
deviating lines of sight have been reported (Nummelin et al. 2001),
the CO$_2$ abundance correlates best of all ice bands with H$_2$O,
revealing important information on the formation history of these
species (\S 6.1). In contrast, it is now well established that the
CH$_3$OH abundance varies by an order of magnitude between different
sight-lines (Dartois et al. 1999; Pontoppidan et al. 2003b). Large
abundance variations are also seen for CO and the species responsible
for the 4.62 and 6.85 \mum\ bands (likely OCN$^-$ and possibly
NH$_4^+$), and weak evidence for NH$_3$ variations exists as well (\S
2.2). No correlation between these variations is apparent, suggesting
that they may have different origins, such as most likely evaporation
of CO, energetic production of OCN$^-$ and possibly a different atomic
H density at the time of CH$_3$OH production.  For the other, low
abundance, species listed in Fig. 4 and further discussed in \S 2.4,
insufficient lines of sight have been observed to investigate
abundance variations as a function of environment.  Finally, no
distinction between the abundances toward low and high mass protostars
is observed. However, the low OCN$^-$ and CH$_3$OH abundance upper
limits for quiescent clouds (\S 4.3) do indicate that ices toward
protostars are qualitatively different.

\begin{figure}
\plotfiddle{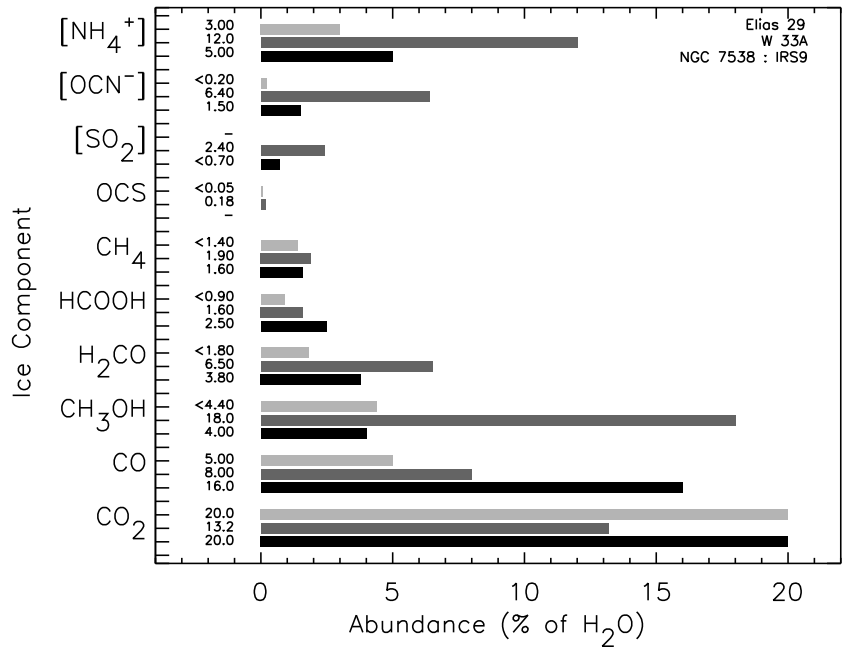}{220pt}{90}{60}{60}{250pt}{-50pt}
\caption{Abundances of molecules in interstellar ices as a percentage
of solid H$_2$O. The identification of molecules in square brackets is
uncertain.  For each molecule the abundance in three lines of sight is
given: NGC 7538 : IRS9 (black), W 33A (dark grey), and Elias 29 (light
grey). Note the large variations of CO, CH$_3$OH, and OCN$^-$
abundances and the much more stable CO$_2$ abundance as a function of
sight-line. Solid H$_2$O column densities are taken from Whittet et
al. (1996; 8$\times$10$^{18}$ cm$^{-2}$), Gibb et al. (2000;
11$\times$10$^{18}$ cm$^{-2}$), and Boogert et al. (2000a;
3.4$\times$10$^{18}$ cm$^{-2}$) for these respective sources.}
\end{figure}

\subsection{Pushing the Confusion Limit: the Identification of Solid NH$\bf _3$}

Despite the wide presence of ammonia in the gas phase interstellar
medium (e.g. Benson \& Myers 1989), detection of this species in ices
is still a much debated issue. Early attempts at detecting the 2.96
\mum\ N-H stretching mode led to abundance upper limits of typically
8\% with respect to H$_2$O (e.g. Knacke et al. 1982; Smith et
al. 1989; Whittet et al. 1996).  A relative abundance of 20-30\% was
claimed toward the Galactic Center source SgrA* (Chiar et al. 2000).
Definite detection of the N-H stretch band is complicated because of
its blend with the O-H stretch mode of H$_2$O, and crystalline
substructures in it (e.g. Smith et al. 1989; Dartois \& d'Hendecourt
2001).  Other vibrational modes of NH$_3$ suffer from similar
complications.  The N-H bending mode at 6.0 \mum\ is blended with the
O-H bending mode of H$_2$O (e.g. Keane et al. 2001). The NH$_3$
umbrella mode at 9.0 \mum\ is blended with the deep absorption band of
silicates. Nevertheless distinct absorption features have been
reported (Lacy et al. 1998; Gibb et al. 2000), leading to abundances
of 10-15\% with respect to H$_2$O.  On the other hand, such an
abundance is inconsistent with the absence of the combination mode at
2.2 \mum\ in the same line of sight (W 33A; Taban et al. 2003), giving
an upper limit of 5\% only.  It must be emphasized that the 2.2 \mum\
spectral region is complex and further study is required to validate
this important conclusion. Finally, ice mixtures of H$_2$O and NH$_3$
produce NH$_3$.H$_2$O hydrates that absorb quite strongly in the
3.3-4.0 \mum\ region. The analysis is again ambiguous because of
blending with the prominent red wing of the 3.07 \mum\ H$_2$O band,
which may be due in part to scattering by large grains (e.g. Hagen et
al. 1983). However, evidence has been presented that a distinct
substructure in this same region, the interstellar 3.47 \mum\
absorption feature, is due to ammonia hydrates as well (Dartois et
al. 2001). This is reinforced by the good correlation of the depth of
this feature with the H$_2$O ice column (Brooke et
al. 1996)\footnote{Note that the 3.47 \mum\ absorption feature in
dense clouds should not be confused with a feature at similar
wavelength, but with a more structured profile, observed in the
diffuse medium and attributed to aliphatic hydrocarbons (Pendleton \&
Allamandola 2002)}. An upper limit of 7\% of NH$_3$ in the hydrate
relative to H$_2$O was derived.  In summary, the various absorption
bands suggest an abundance of solid NH$_3$ in the interstellar medium
of at most 10\% of H$_2$O. Higher abundances may be present in
different sight-lines, in particular toward the Galactic center.
Further study is needed to verify this.

In more general terms, studies of the NH$_3$ molecule illustrate that
often the detection of weak absorption features is limited by spectral
confusion rather than by S/N. Another notorious example is the 6.0
\mum\ absorption band which is only partly explained by the O-H
bending mode of H$_2$O and has additional contributions from likely
HCOOH, H$_2$CO and NH$_3$ (Schutte et al. 1996; Keane et al. 2001) and
perhaps more complex species resulting from heavy processing (Gibb \&
Whittet 2002).

\subsection{Progress in Identifying the 6.85 $\bf \mu m$ Band}

Still no consensus exists on the origin of the prominent 6.85 \mum\
absorption band, discovered in 1979 (Puetter et al. 1979). This band
was observed at low spectral resolution toward a number of massive
protostars with the KAO airborne observatory (Tielens et al. 1984). It
is situated in the spectral region of O-H and C-H bending mode
transitions, and in particular CH$_3$OH has an absorption band at this
wavelength that roughly matches these low resolution spectra. Higher
resolution ISO/SWS data indicate a peak position and smoothness (no
substructures) of the interstellar 6.85 \mum\ band inconsistent with
solid methanol.  Indeed, other bands of this species indicate that
only a fraction ($\sim$10\%) of the 6.85 \mum\ band can be attributed
to CH$_3$OH.  Earlier claims that different bands give different
columns due to scattering effects are not sustained by observations of
other multi-band species (e.g. CO$_2$; Gerakines et al. 1999).

\begin{figure}[t]
\plotfiddle{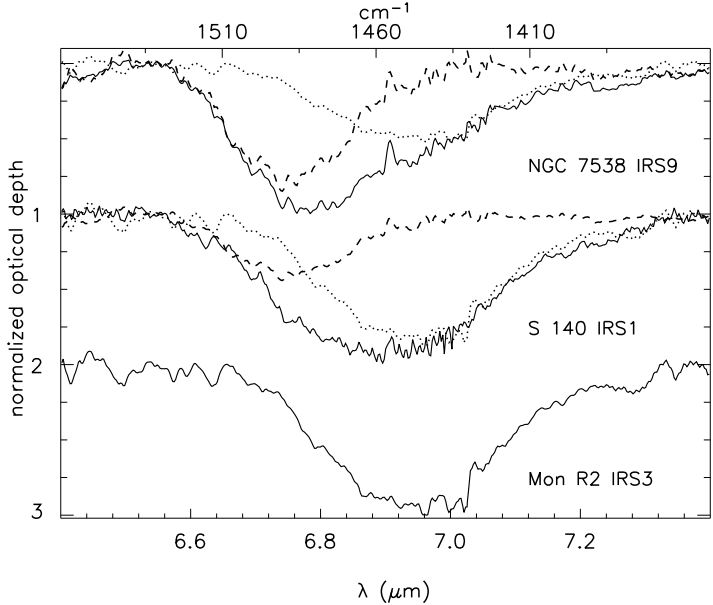}{200pt}{90}{50}{50}{200pt}{-45pt}
\caption{Variation in the 6.85 \mum\ absorption band toward massive
protostars. Dashed and dotted lines indicate the two components with
which all 6.85 \mum\ bands can be fitted (Keane et al. 2001).  The
sources are ordered from top to bottom with warmer line of sight
conditions. The variation of the 6.85 \mum\ profile may well be a
temperature effect.}
\end{figure}

Exhaustive searches for carriers of the 6.85 \mum\ absorption band
have not led to secure identifications (Keane et
al. 2001). Nevertheless, the ISO observations gave significant new
insights into the nature of the carrier.  Large shifts in the peak
position were observed in a sample of lines of sight toward high mass
protostars (Fig. 5).  The 6.85 \mum\ band can be explained by a
combination of two absorption features, with fixed peak position and
width, but varying optical depth. Moreover, the long wavelength
feature becomes relatively deeper for lines of sight with warmer and
more processed ices (\S 4.1; cf. compare Figs. 5 and 7). The
species responsible for the latter band must therefore be less
volatile than the species absorbing at shorter wavelengths. In a
different scenario, a single species could be responsible for both
bands, but it must follow these spectroscopic constraints upon
warm-up.

With this in mind, the NH$_4^+$ proposal of Schutte \& Khanna (2003)
is noteworthy. The red-shift of the 6.85 \mum\ band as a function of
temperature processing is reproduced by heating an NH$_4^+$-containing
laboratory ice.  These laboratory ices have additional bands at 3.25
and 3.47 \mum, where features are observed in the dense interstellar
medium as well (e.g. Sellgren et al. 1994; Brooke et
al. 1996). Despite these successes, the match to the interstellar
features is only approximate and further study is required. In more
general terms, ion production in interstellar ices requires a specific
chemistry. For example, NH$_4^+$ is produced by warm up of NH$_3$ and
acids, which are either deposited (HNCO, HCOOH) or produced by
photo-chemistry. In this, the apparently limited availability of
NH$_3$ in the ices (\S 2.2) and the presence of the 6.85 \mum\ feature
in cold, unprocessed sight-lines seem problematic.

\subsection{Status of Identifying Weak Ice Components}

A set of intriguing absorption features was discovered in the hard to
observe 7-8 \mum\ spectral region: 7.25, 7.41, 7.6, and 7.67
\mum. They are weak and thus so far only detected in a limited sample
of the most deeply embedded objects. The 7.25 and 7.41 \mum\ bands are
attributed to formic acid (HCOOH), and the chemically related formate
ion (HCOO$^-$) respectively (Schutte et al. 1999). Of these, HCOOH is
a more secure identification, because of its contribution to the 6.0
\mum\ band (Schutte et al. 1996). A similar feature at 7.25 \mum\
detected toward the Galactic Center was attributed to aliphatic
hydrocarbons in the diffuse interstellar medium (Chiar et
al. 2000). This alternative identification is consistent with the 3.4
and 6.85 \mum\ bands which have quite different shapes compared to
features at similar wavelengths in dense clouds.  The 7.67 \mum\
absorption band is well explained by the deformation mode of solid
CH$_4$ (Lacy et al. 1991; Boogert et al. 1996). Its identification is
further supported by the presence of gas phase CH$_4$ lines in the
same wavelength region (Boogert et al. 1998; Dartois et al. 1998),
good laboratory fits, and the recent detection of the 3.32 \mum\
CH$_4$ stretching mode (Boogert et al. 2003). The identification of an
additional feature at 7.60 \mum\ is uncertain. It roughly fits with
solid SO$_2$, but detailed laboratory fits are not compelling (Boogert
et al. 1996; 1997). Its abundance upper limit is a factor $\sim$10
larger than that of solid OCS, a value still consistent with grain
surface chemistry models (Palumbo et al. 1997).  However, otherwise
sulphur containing molecules are remarkably under-abundant in
interstellar ices.  A weak feature at 3.9 \mum\ was attributed to
H$_2$S at first (Geballe et al. 1985), but ascribed to CH$_3$OH later
(Allamandola et al. 1992).

Nitrogen containing species are also sparse in the ices.  The 4.62
\mum\ band is attributed to a CN bearing species (likely OCN$^-$;
e.g. Novozamsky et al. 2001), but consumes only a small fraction of
the available N.  NH$_3$ also does not appear to be very abundant (\S
2.2).  Searches for the ``forbidden'' stretch mode of N$_2$ at 4.295
\mum\ lead to upper limits that start to become interesting (Sandford
et al. 2001). If N$_2$ is mixed with the apolar CO phase, the extreme
sensitivity of the \thirteenco\ band to ice composition may improve
further on this (Boogert et al. 2002a; Pontoppidan et al. 2003a).

Similar searches for the ``forbidden'' mode of solid O$_2$ at 6.45
\mum\ and its effect on the CO band profile did provide useful upper
limits (Vandenbussche et al. 1999). Solid O$_2$ consumes less than 6\%
of the O budget. Likewise, the most abundant molecule in space, H$_2$,
is observable in the ices by a weak infrared active band at 2.415
\mum. A detection was claimed toward one source at three times the
solid CO abundance (Sandford et al. 1993), but this needs confirmation
because this is also the region of strong CO band heads (e.g. Taban et
al. 2003). A secure detection of H$_2$ in interstellar ices would have
major implications, because it generally is thought to be too volatile
to freeze out on grains.  Finally, valuable upper limits were also
derived for ethane, ethanol, and hydrogen peroxide (Boudin et
al. 1998).

\section{The Road toward Understanding the Evolution of Interstellar Ices}

In the formation and evolution of interstellar ices a large number of
micro and macroscopic processes can potentially play important roles
(\S 1).  How can observations of interstellar ice absorption bands
help in tracing the importance of these processes?  Before embarking
on specific examples (\S\S 4 and 5), we will discuss the tools that
are available for this purpose.

\subsection{Laboratory Simulations}

In order to identify solid state species in space, to measure their
column densities and to define their physical and chemical properties,
laboratory simulations of space conditions have proven to be very
effective (e.g. Ehrenfreund \& Fraser 2003).  Laboratory simulations
of low temperature ices are achieved by condensing ices as pure gas or
gas mixtures in a high-vacuum chamber on the surface of a
caesiumiodide (CsI) window, cooled by a closed cycle He refrigerator
to 10 K. Infrared transmission spectra are usually obtained with a
Fourier transform spectrometer at a resolution of 0.5-1
cm$^{-1}$. Thermal and energetic processing can be simulated by
stepwise annealing and microwave-excited hydrogen flow lamps,
respectively.

Strong shifts in the peak position, band width and detailed profile
occur as a result of dipole interactions between molecules (e.g.
compare polar and apolar ices; Sandford et al. 1988; \S 1).
Furthermore, the formation of complexes in ice mixtures, e.g. between
CO$_2$ and CH$_3$OH (Ehrenfreund et al. 1998) or between H$_2$O and
NH$_3$ (Knacke et al. 1982), was observed to lead to clear
spectroscopic signatures. When ices convert from the amorphous to the
crystalline state, the crystalline features appear sharper and
red-shifted (Fig. 7). For a realistic comparison of interstellar and
laboratory ices, the effects of grain size and shape need to be taken
into account for the strongest transitions (\S 5). For this, accurate
optical constants are mandatory; these can be measured in the
laboratory (Ehrenfreund et al. 1997). Also, the integrated absorption
cross sections are measured in the laboratory and have been refined
over the last two decades for the most important species
(e.g. Gerakines et al. 1995, Kerkhof et al. 1999). By measuring the
band intensity of an absorption species in space and dividing it by
the integrated absorption cross section measured in the laboratory,
the column density of the given ice species in space can be measured
to a very high accuracy. Finally, photolysis, radiolysis, and surface
chemistry give insights into the basic processes of ice mantle
formation, adsorption, diffusion, reaction efficiency and desorption
(e.g. Gerakines et al. 1996; Hudson \& Moore 1999; Collings et
al. 2003; Hiraoka et al. 1998; Watanabe et al. 2003; \S 6).

\subsection{Where are the Ices?}

Laboratory study is a valuable tool in determining the formation and
processing history of ices. However, understanding the results in an
astrophysical context requires knowledge of line of sight conditions.
The basic question as to where the ices are located along the pencil
beam absorption line of sight is often not easy to answer.  On a large
scale, ices are present in dense molecular clouds. Within these
clouds, different physical conditions and processing histories may
exist in the clumpy cloud interior and at the cloud edges.  The cloud
may also be heated from the interior by shocks and radiation from
protostars.  In the absorption spectrum, features from ices in
protostellar environments may be blended with ice features in
foreground clouds. In turn, ices can be present in a variety of
locations in both protostellar envelopes and disks. Within the disk,
the distance from the disk surface and from the star will influence
ice processing. In all these environments, knowledge of the three
dimensional structure and orientation with respect to the line of
sight is crucial.  Infrared and mm-wave imaging can constrain this
parameter space. Such combined geometry-ice band studies are complex
and have been done for a few sources only. Ices are included in
detailed circumbinary envelope models of L1551 : IRS5 in Taurus,
combining information over all wavelengths (Osorio et al. 2003).
Another example is the Class I protostar Elias 29 in the $\rho$ Oph
cloud (Boogert et al. 2002c). Single dish mm-wave observations reveal
a large column (corresponding to $A_{\rm V}$=20) of foreground
material which may harbor most of the ices in this line of sight
(Fig. 6). High spatial resolution mm-interferometer observations
reveal that some of the ices could be present in an envelope as well,
although detailed SED fits imply that this is a rather face-on system
and therefore the ice contribution from the envelope and disk is
small. Other specific applications of line of sight studies are ices
in the envelopes of high mass protostars (\S 4.1), ices in the edge-on
disks of low mass protostars (\S 4.2), and the special case of ices
toward main sequence or evolved stars behind dense clouds, tracing the
quiescent cloud medium (\S 4.3; Whittet et al. 1998).

\begin{figure}
\plotfiddle{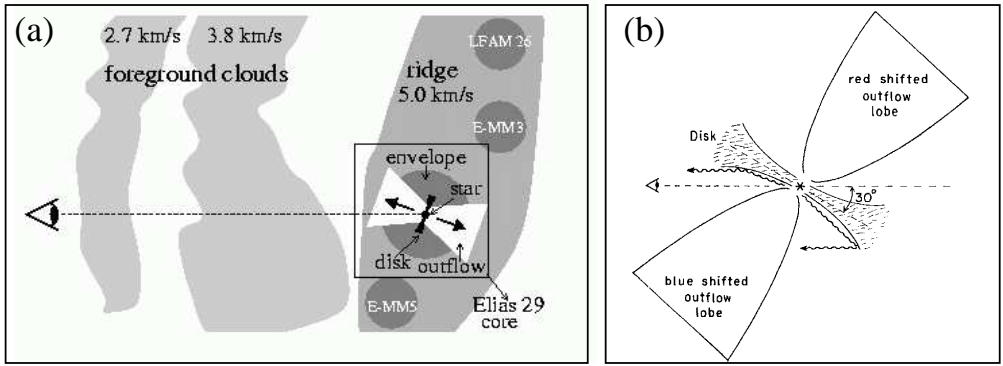}{150pt}{0}{75}{75}{-190pt}{0pt}
\caption{(a): Sketch of the various sources of ice absorption along
the line of sight toward the Class I protostar Elias 29 in $\rho$ Oph,
as found my mm-wave single dish and interferometer observations
(Boogert et al. 2002c). (b): Sketch of the line of sight toward the low
mass protostar RNO 91. The heavily processed ices are claimed to
originate in the inner disk regions (Weintraub et
al. 1994).}
\end{figure}

\section{Ices in Different Environments}

Valuable information on the composition, formation, and processing
history of ices is contained in the shapes of the vibrational mode
absorption bands. Here we discuss band profile studies in different
classes of sight-lines.

\subsection{Ices in the Envelopes around High Mass Protostars}

\begin{figure}
\plotfiddle{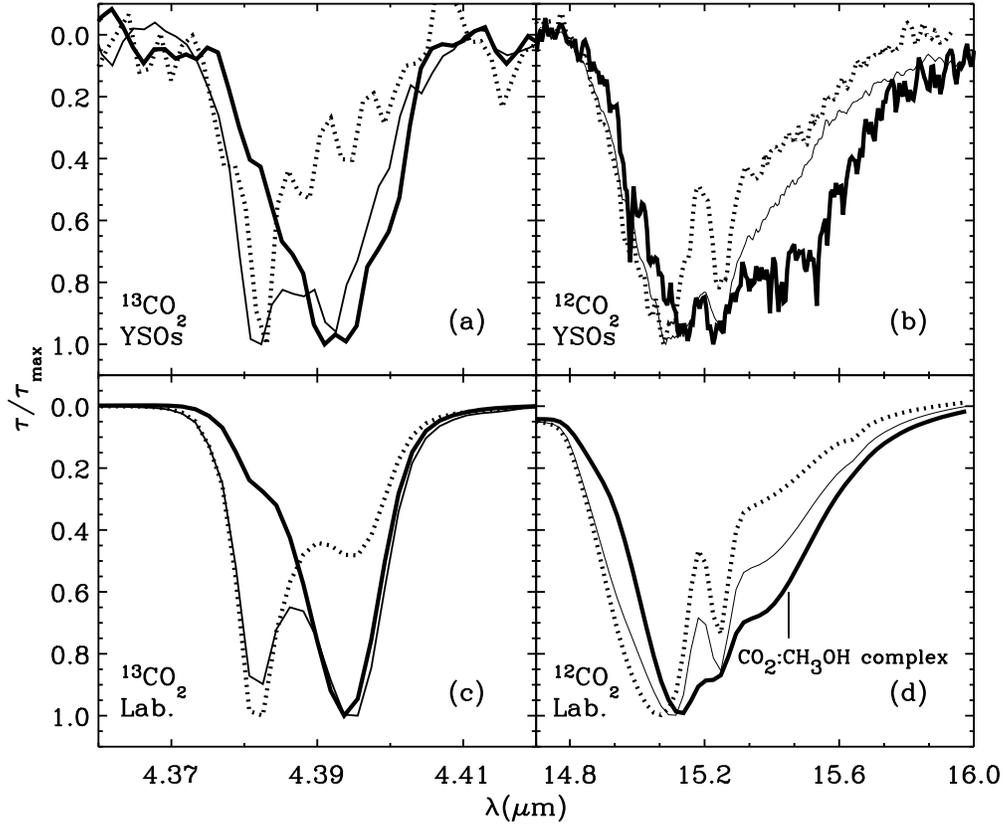}{240pt}{90}{70}{70}{290pt}{-40pt}
\caption{The $\rm ^{13}CO_2$ (a) and $\rm ^{12}CO_2$ (b) bands
observed toward three protostars surrounded by ices with different
thermal histories and chemical compositions.  The bands of S 140 :
IRS1 (dotted line) show the characteristic narrow signatures of
crystalline ices. The envelope of NGC 7538 : IRS9 (thin line) is
clearly much cooler, and presumably younger (Gerakines et al. 1999;
Boogert et al. 2000b). W 33A (thick line) is even cooler and shows a
more prominent red wing due to CO$_2$:CH$_3$OH complexes.  Panels (c)
and (d) show the corresponding laboratory fits to these bands, using a
mixture of H$_2$O, CH$_3$OH, and CO$_2$ in equal quantities but at
different temperatures. A 'CDE' grain shape is assumed.  These plots
show that the $\rm CO_2$ bands are powerful tracers of both the
thermal history and the composition of interstellar ices. The data in
panels (a) and (b) were obtained with the ISO/SWS spectrometer. The
$\rm ^{12}CO_2$ bending mode is a prime target for the SIRTF/IRS
spectrometer.}
\end{figure}

The massive envelopes around high mass protostars have been the focus
of ice studies since the ice absorption bands were discovered (Merrill
et al. 1976; Knacke et al. 1982; Tielens et al. 1984).  These objects
are relatively bright in the near-infrared; the visual extinction
through the envelopes is very high ($A_{\rm V}>100$), and thus their
deep ice bands are easily observable.  The knowledge gained from ice
band observations toward high mass protostars serves as a template
against which the bands in other classes of sight-lines can be
compared.  It was recognized that variations in the profile of the 3.0
\mum\ band of H$_2$O can be attributed to varying degrees of
crystallization due to heating in the different sight-lines (Smith et
al. 1989). Varying abundance ratios of CO in polar and CO in more
volatile apolar ices may well be attributed to evaporation in some
sight-lines (\S 1; Tielens et al. 1991). These early indicators of the
importance of heating effects on ices were strongly confirmed by
observations of the CO$_2$ ice bands by the ISO satellite (Gerakines
et al. 1999).  Sharp substructures in the bands were related to heated
H$_2$O:CO$_2$:CH$_3$OH ices. In these laboratory ices, crystalline
H$_2$O structures are formed and the interaction with CO$_2$ is
reduced, resulting in narrow substructures (Fig. 7). The combined
evidence for ice crystallization, increased gas/solid state ratios,
far-infrared color temperatures and detailed mm-wave gas phase
observations and models led to the conclusion that the envelopes
around massive protostars are progressively heated as a function of
time (Boogert et al. 2000b; van der Tak et al. 2000).  The evidence
for processing is so strong that it has predictive power.  For
example, distinct variations in the so far unidentified prominent 6.85
\mum\ absorption band can be related to this processing sequence (\S
2.3). Also, in future observations of large sets of massive
protostars, the ice bands observed in their envelopes may be used as
{\it tracers} of the evolutionary stage of the protostar. Finally, the
knowledge gained from ice band studies toward massive protostars can
be applied to low mass protostars (\S 7).

\subsection{Solid CO toward Low Mass Protostars}

Observations of ices toward low mass protostars are of special
interest. They allow a look back in time to the formation and
processing history of solar system ices.  The 4.67 \mum\ band of solid
CO is the best studied ice band toward low mass protostars.  It was
detected toward objects in the Taurus (Whittet et al. 1989; Chiar et
al. 1995; Teixeira et al. 1998), $\rho$ Oph (Kerr et al. 1993), Corona
Australis (Chiar et al. 1998), and Serpens (Chiar et al. 1994)
molecular clouds. A recent, comprehensive M band survey of low mass
protostars is presented in Pontoppidan et al. (2003a). Signatures of
processing, such as varying ratios of volatile apolar versus less
volatile polar CO ices (\S 1) were seen in these surveys.  However,
unlike the high mass protostars discussed in \S 4.1, their
significance is harder to assess because of the often poorly
characterized lines of sight.  The complexity is illustrated in the
case-study of the low mass protostar Elias 29 in the $\rho$ Oph cloud
(\S 3.2; Fig. 6). Its ice spectrum is simple, without signs of any
kind of processing. Indeed, its disk has a face-on orientation and a
large fraction of the ices is likely not directly related to the
protostar, but located in foreground clouds. More interesting is the
case in which the disk is in an inclined or edge-on orientation.

\begin{figure}[t]
\plotfiddle{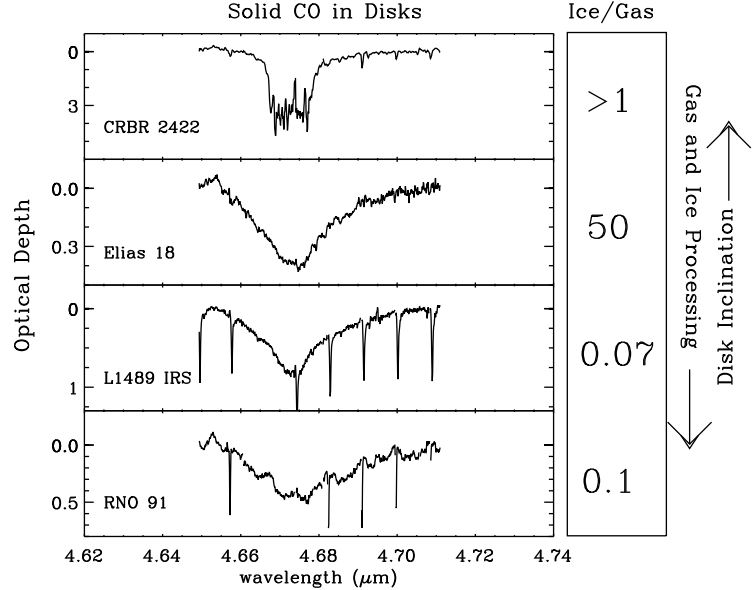}{200pt}{90}{50}{50}{190pt}{-40pt}
\caption{Observations of the CO ice band in edge-on disks, at R=25000
using Keck/NIRSPEC (Boogert \& Blake, in prep.). From top to bottom
the sources are ordered in increasing effects of ice processing,
possible due to decreasing disk inclination (the most edge-on source
on top). Note the differences in optical depth scale and gas phase
line characteristics. Inferred ice/gas ratios are taken from previous
observations (Thi et al. 2002; Shuping et al. 2000; Boogert et
al. 2002b), but estimated for RNO 91.}
\end{figure}

A few claims of the direct detection of ices in edge-on protostellar
disks have been made. Varying degrees of processing are possibly
related to different disk inclinations, tracing ices at different
distances from the star or from the warm flaring disk surface
(Fig. 8).  CRBR 2422.8-3422 has an extreme depletion and the deepest
observed CO ice band observed so far (Thi et al. 2002). Elias 18 has a
very large solid/gas ratio (Shuping et al. 2000).  Warm gas and a much
lower depletion and low apolar/polar ratio are observed toward L1489
IRS (Boogert et al. 2002b).  Finally a highly processed CO band with an
accompanying XCN band is seen in RNO 91 (Weintraub et al. 1994;
Fig. 6).

\subsection{Ices in Quiescent Lines of Sight}

Field stars located behind dense molecular material provide an
invaluable resource for probing the properties of icy mantles in
quiescent regions of molecular clouds (Whittet et al. 1998). Ices
toward such objects are not exposed to the dramatic radiation and
temperature excursions that can occur in the close environment of high
and low mass stars.  Indeed, as expected, and in contrast to many
protostellar sight-lines, the spectra of field stars show large apolar
CO abundances (Chiar et al. 1995), amorphous H$_2$O ice bands (Smith
et al. 1989), and no 4.62 \mum\ OCN$^-$ band (Tegler et al. 1995). The
presence of the 4.27 \mum\ CO$_2$ band as well as the 3.47 \mum\ band
indicates that the energetics of a nearby protostar is not necessary
to produce the carrier of these features (Chiar et al. 1996; Whittet
et al. 1998). This does however not exclude molecule production by
cosmic rays, which can penetrate deep into the cloud (Whittet et
al. 1998).  The meaning of the absence of CH$_3$OH in the direction of
field stars (Chiar et al. 1996) is hard to assess because of the
limited number of observed field stars and the absence of CH$_3$OH in
many protostellar lines of sight as well. A significant limitation of
studying ices toward field stars is the presence of a multitude of
photospheric features in the spectra of these often evolved stars.
While the analysis of ice bands often gains from increased spectral
resolution, this is not the case for the very complex spectra of these
field stars. The well known examples Elias 16 behind the Taurus cloud
and CK 2 behind Serpens are both K giants.  However, some early type
main sequence stars are known as well, e.g.  Elias 25 (Shenoy et
al. 2003; Shuping et al. 2000). This line of sight intersects the edge
of a molecular cloud, with a well known UV field from a nearby star
and allows to study the effect of UV processing.

\section{Isotopes in Ices and the Shape of Interstellar Grains}

Highly valuable information can be obtained from the band profiles of
isotopes in the ices. Weak bands of $^{13}$CO$_2$ (de Graauw et
al. 1996; d'Hendecourt et al. 1996) and $^{13}$CO (Boogert et
al. 2002a) are the only signatures of isotopes detected in interstellar
ices so far. These bands are of particular interest because they give
crucial additional constraints to the composition of ice mantles. For
example, a complication in comparing interstellar and laboratory
spectra is the effect of interstellar grain shape on the band profile.
The profile of strong absorption bands changes drastically as a result
of resonances between the molecular dipoles and ice charge
polarization induced by external radiation fields (e.g. Bohren \&
Huffman 1982). Thus the absorption band profiles of $^{12}$CO and
$^{12}$CO$_2$ are highly dependent on the adopted grain shape if their
concentration in the ice is larger than $\sim$30\%. Due to the large
isotope ratios ($^{12}$C/$^{13}$C=50-100; Wilson \& Rood 1994; Boogert
et al. 2000b) the $^{13}$C equivalents are always strongly diluted in
the ices and particle shape effects are negligible. Laboratory spectra
can therefore directly be compared with the interstellar
observations. Toward the massive protostar NGC 7538 : IRS9 it was
found that both $^{12}$CO and $^{13}$CO trace surprisingly pure CO
ices, a conclusion that is now confirmed in many other lines of sight
(Pontoppidan et al. 2003a). The peak position and width of the
$^{12}$CO band require the icy grains to have a `CDE-type' (continuous
distribution of ellipsoids) grain shape distribution. This may well
represent irregularly shaped grains. Indeed, a similar grain shape is
required in order to fit the interstellar $^{12}$CO$_2$ and
$^{13}$CO$_2$ band profiles (Gerakines et al. 1999; Boogert et
al. 2000b).

Finally, the band of a third isotope, crystalline HDO at 4.1 \mum, was
claimed in interstellar spectra (Teixeira et al. 2000). In a
re-analysis of the data, this feature was considered to be an artifact
in the NGC 7538 : IRS9 line of sight, but confirmed in others (Dartois
et al. 2003).  However, this feature originates from CH$_3$OH and not
HDO. HDO remains undetected in low mass protostars as well, with
significant abundance upper limits (Parise et al. 2003).

\section{Tracing Chemistry on Dust Surfaces}

Physical processes, such as evaporation and crystallization, lead to
clear spectroscopic signatures that indeed are observed toward
protostars (\S 4).  What is the observational evidence for chemical
processes, such as formation of molecules on cold grain surfaces or by
energetic reactions induced by UV photons or cosmic rays?

\subsection{Grain Surface Chemistry}

Astrochemical theories indicate that only a few classes of
low-temperature grain-surface chemical reactions are necessary to form
most of the molecules observed in ices (e.g., Allen \& Robinson 1977;
Tielens \& Charnley 1997; Herbst 2000). Fully stochastic treatments to
model surface diffusion and surface reactions are commonly applied
(e.g. Charnley 2001). Water (H$_2$O), methane (CH$_4$), ammonia
(NH$_3$), and hydrogen sulphide (H$_2$S) are readily produced by
exothermic H additions to O, C, N, and S atoms. Indeed, the high
H$_2$O ice abundance in molecular clouds ($\sim 10^{-4}$ with respect
to $N_{\rm H}$) can only be produced by such efficient grain surface
reactions. The abundance of other hydrogenated species is however low,
indicating that the presently observed ices originate from a phase
with relatively low atomic and high molecular abundances.  For
example, the solid CH$_4$ column is only 1\% of that of H$_2$O and at
the time of its formation most of the C must have been locked up in
CO. The origin of CH$_4$ on grains, like H$_2$O, is supported by its
low gas/solid ratio (Boogert et al. 1998) and its presence in the
polar H$_2$O--rich ice phase. Similarly, the low NH$_3$ abundance (\S
2.2) may well indicate that at the time of grain mantle formation most
of the N is locked up in N$_2$.

Within the same picture, hydrogenation of CO was suggested (e.g.,
Tielens \& Charnley 1997; Caselli et al. 2002) to be the source of the
solid methanol seen in many lines of sight towards protostars
(CH$_3$OH/H$_2$O=5--25\%; Dartois et al. 1999; Pontopiddan et
al. 2003b). Indeed, the intimate mix of solid CH$_3$OH with H$_2$O and
CO$_2$ required to explain the CO$_2$ bending mode (Fig. 7) supports
this.  In this picture, the large variations of solid CH$_3$OH
abundances in different sight-lines reflect variations in the atomic H
abundances at the time of CH$_3$OH formation. However, the grain
surface origin of CH$_3$OH is not fully established. For example, its
absence in many lines of sight may indicate that specific conditions,
and perhaps chemical mechanisms other than grain surface chemistry are
required for CH$_3$OH production.  An independent indicator of its
origin comes from the D/H ratio.  Both methanol and formaldehyde
(H$_2$CO) observed in protostellar cores have large D/H ratios,
indicating formation of these species at low temperature ($<20$ K;
Loinard et al. 2000; Parise et al. 2002).  Low temperature gas phase
chemistry can however also be responsible (Roberts \& Millar 2000;
Ceccarelli et al. 2002), and currently CH$_3$OH formation is not fully
understood.  Finally, initial experimental reports that hydrogenation
of CO could produce methanol at low temperatures (Hiraoka et al. 1994;
1998) are strongly supported by Watanabe et al. (2002, 2003) and
discarded by other studies (Hiraoka et al. 2002).

While the light H and D atoms can saturate CO by tunneling through the
activation barrier (Tielens \& Hagen 1982; Tielens 1983), the addition
of heavier atoms to a closed-shell molecule such as CO is much
harder. The current knowledge on activation barriers is scarce, and
thus this pathway is not well established.  Only recently atomic
reactions on analogue surfaces have been studied in detail (e.g.,
Hiraoka et al.  1998; Pirronello et al. 1999). However, it was
realized that in interstellar space even a reaction barrier does not
inhibit heavy atoms to react with CO. This is because reactions in
space are diffusion rather than reaction limited (Tielens \& Charnley
1997). Thus appreciable amounts of CO$_2$ can be formed by O addition
to CO on cold grain surfaces (Tielens \& Hagen 1982).

A separate way to trace chemical history is by measuring isotope
ratios. In one line of sight $^{12}$CO/$^{13}$CO ratios of 71
(3$\sigma$) and $\sim$77 were deduced in the solid and gas phase
respectively, comparable to the solid $^{12}$CO$_2$/$^{13}$CO$_2$
ratio of 80 (Boogert et al. 2002a). This indicates that CO$_2$
originates from oxidation reactions of existing CO rather than from
gas phase C, which would have almost certainly led to large
$^{12}$CO$_2$/$^{13}$CO$_2$ fractionation.

Finally, while grain surface chemistry can efficiently produce many
molecules, it was of small influence during the formation of apolar
ices. Recent high resolution studies of solid CO have provided new
insights into the nature of apolar ices (Boogert et al. 2002a, 2002b;
Pontoppidan et al. 2003a). Most remarkable is the great similarity of
the band profiles in many lines of sight and the purity of the CO ice
in all these environments (\S 5). At a particular stage in the
molecular cloud evolution simple CO ices are formed rather than
complex mixtures with other apolar species such as O$_2$ or CO$_2$
(`dirty' ices). Presumably at this stage all oxygen is depleted
already. The conditions and time scale at which pure CO ice is formed
may be traced back in the band profile.  Most sight-lines show besides
the `classic' narrow central peak and broad red wing of apolar and
polar CO components (\S 1; \S 4.2), a newly discovered blue wing as
well.  This wing, visible in the presence of linearly polarized light,
may indicate that CO is present in a largely crystalline phase,
reflecting its formation history (Pontoppidan et al. 2003a).

\subsection{Energetic Processing}

Interstellar ices are potentially subject to processing by photons and
particles in a wide energy range.  In marked contrast to the amount of
effort expended in trying to model surface reactions, there has been
almost no theoretical modeling of the kinetics associated with the
photolysis or radiolysis of the bulk ice mantle (however, see Ruffle
\& Herbst 2001; Woon 2002). In contrast, energetic processing of ices
has historically enjoyed the attention of experimental groups.

Simply warming of ices initially formed by cold grain surface
chemistry is sufficient to significantly alter the chemical
composition. For example, CO$_2$ is efficiently produced in CO:H$_2$O
mixtures once warmed enough to surpass the small activation energy
barrier needed to oxidize CO to form CO$_2$ (e.g. Roser et al. 2001;
however, see \S 6.1). Heating of ice mixtures can also result in
complex molecule formation. For example polymerization of H$_2$CO in a
mixture with NH$_3$ forms polyoxymethylene (POM; (-CH$_2$-O-)n;
Schutte et al. 1993).  Finally, the ion NH$_4^+$ is produced after
heating the acid/base mixture of NH$_3$ and HCOOH (Schutte \& Khanna
2003; \S 2.2).

More energetic forms of processing, such as UV photolysis (e.g.,
Gerakines et al. 1996) and proton irradiation (radiolysis; e.g.  Moore
\& Hudson 1998) of bulk ices, have been extensively studied in the
laboratory. These efficiently produce radicals, leading to complex
species.  In principle cosmic rays can have a significant influence on
interstellar ices, because they can penetrate through clouds and ice
surfaces and thus modify structure and composition of icy particles,
including sputtering of the particle surfaces (Cottin et al. 2001).
For example, radiolysis of H$_2$O/CO ice mixtures can produce high
abundances of H$_2$CO and CH$_3$OH (Hudson \& Moore 1999), although
HCOOH is over-produced relative to the HCOOH/CH$_3$OH observed in
interstellar ices.  Radiolysis of ice mixtures containing CO and
C$_2$H$_2$ also produces ethane, as well as several other putative
mantle molecules such as CH$_3$CHO and C$_2$H$_5$OH (Hudson \& Moore
1997; Moore \& Hudson 1998).  A main concern of these experiments is
that more species are produced than can be accounted for in the
interstellar ices.

One interstellar molecule that does likely originate in energetically
processed ices is the carrier of the so-called 4.62 \mum\ `XCN'
absorption feature, observed in many sources (e.g., Tegler et
al. 1995; Pendleton et al. 1999).  Laboratory experiments indicate
that `XCN' is produced both by photolysis (Lacy et al. 1984; Bernstein
et al. 2000) and by radiolysis (Hudson \& Moore 2000; Palumbo et al.
2000).  At present, the best candidate for the identity of this
carrier appears to be the OCN$^-$ ion, formed in a solid state
acid-base reaction between NH$_3$ and isocyanic acid (HNCO; see
Novozamsky et al. 2001 and references therein).

The identification of the 4.62 \mum\ band as a result of energetic
processing offers the possibility to find other energetically produced
species by correlation studies. Gibb \& Whittet (2000) find that the
excess absorption in the 6.0 \mum\ band, i.e. the fraction not caused
by the H$_2$O bending mode, correlates with the 4.62 \mum\ band
depth. It is argued that organic refractory matter (ORM), the residue
of heavily processed simple ices, is responsible for the 6.0 \mum\
excess absorption, rather than HCOOH and H$_2$CO (\S 2.4; Schutte et
al. 1996). At present, this identification needs further study.
Unfortunately, as argued in \S 2.2, specific detection of complex
species in the ices is hindered by the weakness of the features and
confusion limitations. Such complex species may be better studied
indirectly by their rotational transitions at radio wavelengths, after
they have evaporated from the grains in the warm regions surrounding
protostars.  

Overall we do conclude, however, that energetic processing appears to
be of small influence on the ices.  The spectra of {\it all}
protostars are dominated by the absorption bands of simple species,
that likely have been produced by cold grain surface
reactions. Furthermore, high-mass stars do not show significantly
different ice signatures compared to low-mass stars.  Photolytic
processing of ices apparently only has limited influence on the ice
chemistry due to the strongly attenuated UV irradiation in dense
clouds.

\section{Conclusions and Future}

Increasingly sensitive observations of interstellar ices have revealed
several prominent, and often unexpected, trends, providing clues to
the evolutionary pathway of interstellar to solar system ices. One
such trend is the remarkable simplicity of the interstellar ice
spectrum.  Ices consist mostly of simple species, likely the result of
atomic O and H addition reactions on dust surfaces.  A stable balance
between some of these reactions is reflected in the remarkably
constant solid CO$_2$/H$_2$O abundance ratio (17\%) throughout the
Galaxy and the commonly observed intimate mix of these species with
CH$_3$OH. The solid CO band reflects an entirely different ice
formation environment, but again shows little compositional variation
throughout the Galaxy.  This apolar ice phase is remarkably CO--rich.
The similarity of ice composition in sight-lines sampling a variety of
environments proves that to first order the production of complex
species by energetic processing of simple ices is minor. However, at a
low level energetic processing can presently not be excluded. Perhaps
the variations of CH$_3$OH and `XCN' abundances, and the `6.0 \mum\
excess' (Gibb \& Whittet 2002) are linked to locally increased photon
or cosmic ray fluxes. If organic residues indeed form from ices, the
inefficiency of energetic processing in dense clouds requires that
multiple cycling of grains between the dense and diffuse interstellar
medium takes place (Greenberg et al. 2000). Finally, although
energetic processing is weak, the effects of thermal processing of
ices, i.e. evaporation, crystallization and segregation, are commonly
observed. In fact, ice bands can serve as tracers of the thermal
history of circumstellar environments.

Many key issues on the formation and evolution of interstellar ices
are still poorly known, however. In order to finally understand
interstellar ice evolution a concerted effort is required, involving a
variety of observational and laboratory techniques and related theory
projects. Astronomical spectra sample different grain populations in
one line of sight. Line of sight conditions need to be determined from
complementary information, including radio observations of gas phase
species and imaging of the star-forming region. Particularly exciting
is the prospect of studying the properties of ices in disks that are
well characterized by millimeter-wave interferometry (CARMA, ALMA,
SMA) and infrared ro-vibrational CO line studies (Boogert et
al. 2002b, Najita et al. 2003). Also, study of ice features in lines
of sight with known UV radiation fields will shed more light on the
importance of energetic processing of ices (e.g. Shenoy et al. 2003).

The importance of high spectral resolution in the infrared to separate
weak, blended ice bands and gas and telluric lines, cannot be
overstated. Such instruments are now available on large ground based
telescopes like Keck and VLT. In spectral regions severely hindered by
the earth atmosphere, such as the important 5-8 \mum\ region, high
spectral resolution will be provided by spectrometers mounted on the
airborne SOFIA observatory.  Further in the future the planned JWST
and proposed ABE missions will become guaranteed milestones in the
area of interstellar ices.  For example, these instruments allow an
improved assessment of the significance of the apolar species N$_2$
and O$_2$ in the ices. Also, they will be a major leap forward in the
continued search for the weak signatures of energetic processing.

In the near future, sensitive space based observations will be
possible with the SIRTF/IRS spectrometer.  Its spectral resolution is
limited (Fig. 2), but, in particular for the CO$_2$ bending mode,
sufficient for detailed laboratory analysis tracing thermal processing
and molecular complexity toward large samples of low mass protostars
(Fig. 7). Linking the CO$_2$ analysis with ground based observations
of CH$_3$OH is needed to understand the chemical history of this
molecule and the large abundance fluctuations in different
sight-lines.  This is of particular interest because CH$_3$OH is a
starting point of complex molecule formation.

Finally, continued laboratory work will provide much needed chemical
reaction efficiencies in astrophysically relevant environments.
Laboratory simulations are also necessary to determine the microscopic
structure of interstellar ices, which regulates key processes such as
outgassing (e.g. Jenniskens \& Blake 1996) and coagulation behavior
(e.g. Chokshi et al. 1993). Perhaps related to this is the question
why the interstellar ice bands are best fit with a `CDE'-type shape
distribution (\S 5)? Other major issues that have to be settled in the
laboratory include the presence of PAH molecules and ions in ices
(Gudipati \& Allamandola 2003), the identification of the 6.85 \mum\
band, and the physical interaction between the components of
inhomogeneous ices (Collings et al. 2003).

\acknowledgments

We thank J. V. Keane for kindly providing Fig. 5, G. A. Blake for
collaboration in the production of Fig. 8, T. Y. Brooke and an
anonymous referee for carefully reading this manuscript.  The research
of A.C.A.B. is supported by the SIRTF Legacy Science program and by
the Owens Valley Radio Observatory through NSF AST-9981546.

\end{document}